\documentclass[preprint]{aastex}

\usepackage{graphicx}
\usepackage{txfonts}
\usepackage{mathrsfs}

\shorttitle{MRI: a possible mechanism for funnel flows?}
\shortauthors{S.
Do\u{g}an \and E. R. Pek\"{u}nl\"{u}}

\begin{document}


\title{MRI: a possible mechanism for funnel flows?}


\author{S. Do\u{g}an\altaffilmark{1,2} and E. R. Pek\"{u}nl\"{u}\altaffilmark{1}}
 \affil{$^1$Department of Astronomy and Space Sciences, Faculty of Science, University of Ege, 35100, Bornova, \.Izmir,
Turkey} \affil{$^{2}$Theoretical Astrophysics Group, University of
Leicester, Leicester LE1 7RH, UK}

\email{suzan.dogan@ege.edu.tr}



\begin{abstract}
The magnetorotational instability (MRI) has been suggested to have
an important role on the dynamics of accretion disks. We investigate
MRI as an alternative way for guiding the plasma from the disk to
the funnel flow at the disk-magnetosphere boundary of classical T
Tauri stars (CTTSs) by considering the diamagnetic effects. We solve
the magnetohydrodynamic equations by including the effect of both
the magnetic field gradient and the perpendicular (to the field)
velocity gradient produced by the magnetization current at the
disk-magnetosphere boundary for the first time. \emph{Diamagnetic
current modified MRI} produces a non-propagating mode which may lift
the plasma from the disk towards the vertical magnetic field lines.
Our model also shows that the diamagnetic effects play an important
role in triggering the MRI. The instability becomes more powerful
with the inclusion of the gradient in the magnetic field and the
perpendicular velocity.
\end{abstract}


\keywords{Accretion, accretion disks -- instabilities --  MHD -- plasmas}



\section{Introduction}
One of the unsolved problems related to the accretion of matter onto
a rotating star with a dipole magnetic field is identifying the mechanism that
guides the plasma from the disk into the funnel flow (FF)
(see Ghosh \& Lamb 1978; Pringle \& Rees 1972; Romanova et
al. 2002). The basic picture of the disk-magnetosphere interaction
in magnetized stars is generally described as follows: It is
commonly assumed that the accretion flow is disrupted within the
magnetospheric boundary layer where the magnetic forces become
dominant in determining the motion of the plasma, then the plasma is
funneled onto the polar caps of the star. However, the interaction
between the disk and the magnetosphere is exceedingly complicated.

Over the last few decades, there have been a number of theoretical
efforts to understand the nature of the disk-magnetosphere
interaction in a magnetized star (e.g. Pringle \& Rees 1972; Lamb et
al. 1973; Ghosh \& Lamb 1978; Camenzind 1990; K\"{o}nigl 1991;
Spruit \& Taam 1993; Shu et al. 1994; Lovelace et al. 1995; Li et
al. 1996.). The magnetospheric accretion models were proposed first for
neutron stars and black holes. Then, these models were adopted
and developed for magnetic T Tauri stars. Johns-Krull \& Gafford
(2002) examined three analytic theories assuming that accretion is
controlled by magnetosphere in classical T Tauri stars (CTTSs). They
claim that while the surface magnetic fields of BP Tau and TW Hya
are not dipolar, it is likely that the dipole component dominates at
the inner disk-magnetosphere boundary. They suggest that the disk
truncation radii in CTTSs is in the stellar radius range of
3R$_{*}$ - 6R$_{*}$,where the dipole component governs the
disk-magnetospheric interaction. K\"{u}ker, Henning \& R\"{u}diger
(2003) also studied the disk dipolar magnetic field interaction in
CTTSs. They found the critical field strength for the disk disruption
to lie between 1 and 10 kG. They also consider the
possibility of the drainage at the inner parts of the disk by
magnetically enhanced accretion.

In recent years, several authors presented observational evidences
for the magnetospheric funnels on to CTTSs. Muzerolle et al. (1998) used the model favouring
magnetospheric accretion through dipole magnetic field lines and
successfully explained the observed spectral line shapes. Stempels \& Piskunov
(2002) presented the results of observations done with UVES/VLT on
one of the CTTSs RU Lup. Their main interest was to reveal the
properties and geometry of the accretion process. The authors reported
that the observational results were in agreement with the
magnetospheric accretion model. Stempels \& Piskunov (2002) argue
that magnetic accretion model proved itself to be consistent with
some of the emission lines' radiative transfer calculations.
 Magnetospheric accretion model for neutron stars
 introduced by Ghosh \& Lamb (1979) later applied to T Tauri stars
by Uchida \& Shibata (1985) and K\"{o}nigl (1991) is regarded as
consistent with complex observational picture in CTTSs case by
Stempels \& Piskunov (2002). By using Doppler imaging technique,
Strassmeier et al. (2005) investigated one of the weak-lined T Tauri
star (WTTS), MN Lupi in order to tell the spectral signatures of the
accretion flow from the chromospheric and photospheric magnetic
activity. They obtained Doppler images of hot spots at high stellar
latitude and related it to the accretion shocks produced by the disk
material funneling along the magnetic field lines. Donati et al.
(2011) studied one of the CTTSs, TW Hya, and recently reported the
results of spectropolarimetric observations. They observed a
near-polar region of accretion-powered excess Ca II and He I
emission which coincides with the main magnetic pole of TW Hya and
they claimed that the accretion occurs mostly polewards at the
stellar surface.

Many numerical simulations have also been made of accretion onto a
magnetized star (e.g., Miller \& Stone 1997; Romanova et al. 2002,
2011; Kulkarni \& Romanova 2005; Long et al. 2008). Romanova et al.
(2002) investigated disk accretion onto a rotating magnetized star
and the associated funnel flows by performing a set of simulations
for different stellar magnetic moments and rotation rates. In their
investigation, they found that the dominant force driving plasma
into FF is the pressure gradient force. Although they made numerical
applications to T Tauri stars only, they claimed that their results
are also valid for cataclysmic variables and neutron stars in X-ray
binaries. Romanova et al. (2011) also performed axisymmetric MHD
simulations of accretion onto magnetized stars from
magnetorotational instability (MRI)-driven disks. Close to the star,
they observed that the disk is stopped by the magnetic pressure of
magnetosphere and matter is lifted through a funnel stream.

Plasma entry into magnetospheres is generally explained by plasma
instabilities at the disk-magnetosphere interface in the radial flow
case. A Rayleigh-Taylor instability is expected to occur at the
interface, since the magnetic field may act as a light fluid
supporting a heavy fluid, i.e. plasma  (Elsner \& Lamb 1977, Arons
\& Lea 1976). In addition, the Kelvin-Helmholtz instability is
expected to have a role in plasma entry, because of the relative
motion of the plasma with respect to the magnetosphere (Arons \& Lea
1976). Plasma entry via diffusion, magnetic reconnection and
loss-cone mechanism have also been discussed by Elsner \& Lamb
(1984). Varniere \& Tagger (2002) studied the accretion-ejection
instability in magnetized disks. The authors claim that the
instability can produce slow magnetosonic waves and they expect that
these waves will lift the plasma above disk. Recently, Fu \&
Lai (2012) studied the dynamics of the innermost accretion flows
around compact objects. Their investigation includes a comprehensive
study of the large scale Rayleigh-Taylor and Kelvin-Helmholtz
instabilities associated with the disk-magnetosphere interface of a
rotating magnetized system.

The goal of this paper is to present the role of MRI in the
interaction between the inner disk and the magnetosphere of CTTSs.
Balbus \& Hawley (1991, hereafter BH91) showed that Keplerian disks
with a weak field which fulfills the frozen in condition are
dynamically unstable. If the magnetic field is weak, the
perturbations generate an unstable mode which is non-propagating and
evanescent. The importance of MRI lies in the generality in its
applicability. After BH91 established the importance of MRI
in the dynamics of the accretion disks, this instability has been
received much attention and studied by many investigators over the
last two decades. Most of these authors drew attention on the
necessity of considering the non-ideal MHD effects in
protoplanetary disks (PPDs). For example,Wardle (1999)
found that the Hall effect modifies the growth rate of the
instability. When the Hall current is dominated by the negative
(positive) species, the parallel case becomes less (more) unstable.
Balbus \& Terquem (2001) analysed the Hall effect in
protostellar disks and found that the inclusion of the Hall effect
destabilizes the disk with any differential rotation law.
R\"{u}diger \& Shalybkov (2004) investigated the linear instability
in a magnetic Taylor-Couette (TC) flow with Hall effect.  One of
the major conclusions they had drawn was that while the shear in
disk is negative, the Hall instability combines with the MRI.
Although their main interest was the TC flow they also commented on the
Hall effect on the MRI in astrophysical objects (white dwarfs,
neutron stars and protoplanetary disks).

More recently, Armitage (2011) argued that since the gas in
PPDs is cool, dense and has a very low ionization fraction, one
needs to take into consideration the non-ideal MHD effects, like
Ohmic resistivity, Hall effect and ambipolar diffusion. Besides,
these non-ideal terms are effective at different radial and
vertical distances (z) in the disk. In that part of the
disk where magnetic field is strongly coupled to electrons but not
to the ions, the Hall effect is the most important one (Armitage
2011). He also clearly states that the conductivity and ionization
fraction of the innermost disk fluid are high. Thus, non-ideal MHD
effects play an important role at transporting angular momentum
outward. Under these conditions the disk fluid interacts with the
stellar magnetosphere and MRI is highly likely to set in. One of the
most important conclusions Armitage (2011) draws is the absence of
the non-linear solutions covering the non-ideal conditions, the Hall
effect and the Ohmic and ambipolar diffusion in PPDs. He points to
the future works which are to take into account of the global
effects over long time scales. Besides, Bai (2011)
investigated the MRI-driven accretion in protoplanetary
disks (PPDs) by considering the non-ideal MHD effects
including the Ohmic resistivity, the Hall effect and the ambipolar
diffusion. Bai (2011) also warns the reader about the
necessity of careful exploration of the Hall regime with numerical
simulations which is yet to be done. If the Elsasser number, which
measures the relative importance of the Lorentz force and the
Coriolis force is less than unity, then the non-ideal MHD terms
dominates in the MRI active region and the linear properties of MRI
change  considerably (Bai 2011). In this region, the Hall effect and
the ambipolar diffusion are the dominant ones. In our investigation
we take only the Hall effect into consideration.

In the present investigation we are interested in the diamagnetic
effect and its consequences as the gradient in the magnetic field
and the perpendicular velocity (to the magnetic field) at the
disk-magnetosphere boundary of CTTS. We investigate the instability
of the mode with a wave vector perpendicular to the disk. We improve
a model including the effect of diamagnetism on numerical
growth rates of the unstable mode. The paper is structured as
follows: In the following section, we briefly review the diamagnetic
behaviour of the disk, then we present the mathematical formalism of
diamagnetism. We also present the linearized form of the basic MHD
equations and obtain the general form of the dispersion relation in
section 2. In section 3, a detailed analysis of the effect of the
magnetic field and perpendicular velocity gradients produced by
magnetization on the numerical growth rates of the unstable mode is
carried out. Finally in section 4, we summarize our conclusions from
this work.

\section{Basics}
\subsection{Theoretical background on diamagnetism}
The diamagnetic effect arises when particles moving in an external
magnetic field create their own field. In this investigation, we
assume that the charged particles are frozen into the magnetic field. If
charged particles gyrating around the magnetic field lines produce a net
current at the boundary of a closed circuit, this current in turn
produces a new magnetic field (Singal 1986, Bodo et al. 1992). The
direction of this new and local magnetic field will be the same as
the global magnetic field outside the circuit and the opposite
within the circuit. The net magnetic field inside the region will
therefore be lower than that of outside and a gradient in the magnetic
field will be produced. Electrically charged particles gyrating under the influence of
\textquotedblleft $\nabla B$ \textquotedblright~and \textquotedblleft $\nabla \times \vec{B}$ \textquotedblright~will give rise to a drift
current at the border of the region (see Fig.1).
Because the magnetic moment ($\mu=mv_{\bot}^{2}/2B$), the first
adiabatic invariant is conserved, we expect the magnetic field
gradient to produce a gradient in the perpendicular velocity
($v_{\bot}$) of the particles.

The counterfield produced by particles can be expressed by
magnetization, defined as the magnetic moment per unit volume
(Singal 1986):
\begin{equation}
\emph{\textbf{M}}=\int \limits_{4\pi} \int \limits^\infty \limits_0
N(E,\theta) \mu (E,\theta)\,dE\,d\Omega.
\end{equation}
Here, $N(E,\theta)\,dE\,d\Omega$ is the number density of charged
particles having velocity within $d\Omega$ around pitch angle
$\theta$ and energy within $dE$ around E. By using the definition of the magnetic moment, the magnetization is found as
\begin{equation}
\emph{\textbf{M}}=-\frac {\emph{\textbf{B}}}{3B^{2}}(W_{\rm
r}+2W_{\rm nr})
\end{equation}
where $W_{\rm r}$ and $W_{\rm nr}$ are the energy densities of the
relativistic and non-relativistic particles, respectively (Singal
1986; Bodo et al.1992). In our investigation we discarded the
relativistic electrons flowing in the currents. Therefore, Eq. (2)
can be rewritten in terms of the perpendicular component of the
kinetic energy density of non-relativistic electrons as
\begin{equation}
\emph{\textbf{M}}=-\frac {2\emph{\textbf{B}}}{3B^{2}}W_{\rm k}
\end{equation}
where $W_{\rm k}=nmv^{2}_{\perp}/2$, with \emph{n} the particle density. In the close neighborhood of
the magnetization current carrying circuit, the net magnetic field
may be written as
\begin{equation}
\emph{\textbf{B}}=\emph{\textbf{H}}+4\pi
\emph{\textbf{M}}=\emph{\textbf{H}}-\frac{8\pi}{3}\frac
{\emph{\textbf{B}}}{B^{2}}W_{\rm k}=\emph{\textbf{H}}-\frac
{1}{3}\frac{W_{\rm k}}{W_{\rm
B}}\emph{\textbf{B}}=\emph{\textbf{H}}-\varepsilon \emph{\textbf{B}}
\end{equation}
where $W_{\rm B}=B^{2}/8\pi$ is the magnetic field energy density
and $\varepsilon=W_{\rm k}/3W_{\rm B}$, which is called the
\textquotedblleft magnetization parameter\textquotedblright  in
Devlen \& Pek\"{u}nl\"{u} (2007, hereafter DP07).

In the presence of diamagnetism, the total current density may be
written as
\begin{equation}
\emph{\textbf{J}}=\textbf{\emph{J}}_{\rm ext}+\textbf{\emph{J}}_{\rm
mag}=\frac{c}{4\pi }\nabla \times \emph{\textbf{H}}+c\nabla \times
\emph{\textbf{M}}=\frac{c}{4\pi }\nabla \times \emph{\textbf{H}}-c
\nabla \times \frac {2W_{\rm k}}{3B^2}\textbf{\emph{B}}.
\end{equation}
After some vector operations the current density is found as below
(DP07):
\begin{equation}
\emph{\textbf{J}}=\frac{c}{4\pi } \left[{\rm (}1-\varepsilon {\rm
)}\nabla \times \emph{\textbf{B}}+2\varepsilon \frac{\nabla B}{B}
\times \emph{\textbf{B}}-2\varepsilon \frac{\nabla {\rm v}_{\bot }
}{{\rm v}_{\bot } } \times \emph{\textbf{B}}\right].
\end{equation}

MRI is shown to be the source of turbulence in disks (BH91,
Balbus \& Hawley 1998, hereafter BH98). If fluid elements with outwardly decreasing velocity field
couple with the magnetic field then a torque is produced which
causes enhanced outward angular momentum transport.

In the present study, we intend to seek a solution to the cause of
funnel flow departing from the inner boundary of the disk.
The equilibrium magnetic field at the above mentioned location is assumed to be in the z-direction and perpendicular
to the disk. If the angular velocity of the inner portion of the
differentially rotating disk is more or less equal to that of the
last closed field line of the co-rotating magnetosphere, we may
expect the disk material be trapped at this border. Then, the
positively and the negatively charged particles will acquire drift
velocities in the opposite senses perpendicular to the local
magnetic field due to the curl and the gradient of the dipole magnetic field, the latter of
which is inversely proportional to the third power of the radial
distance. This will bring about a local current flowing at the
border of the inner disk and the co-rotating magnetosphere. This
local current will generate its own magnetic field in such a way as
to increase the magnetic field intensity outside and to decrease it
inside the circuit. So called diamagnetic effect will generate a
magnetic field gradient in the radial direction. The weaker magnetic
field, necessary for the MRI to set in, within the circuit may
trigger the perturbations at the border. If the gradient of the
magnetic field is steep enough then the magnetic pressure force will
be exerted upon the disk material in the negative radial direction.
Assuming that the diamagnetic current lasts long enough to maintain
the magnetic field gradient which through the magnetic pressure
force pushes the disk material into the diamagnetic current
circuit and thus the condition between the sound speed ($c_{\rm s}$) and the  Alfv\'{e}n speed ($v_{\rm A}$), $c_{\rm s}^{2}>v_{\rm A}^{2}$ is fulfilled which is a necessary condition for MRI to set in.

The magnetic field which was in the z-direction at
equilibrium, now, by the push of the frozen-in trapped particles at
the neighbourhood of the border will acquire a curly shape.
Contribution to the drift velocity, originally produced by the
magnetic field gradient, of the electrons from the curvature of the
field lines will enhance the diamagnetic current which in turn
weakens the magnetic field within the circuit. This is an additional
agent to cause MRI to set in. The direction of the electron drift
velocity is in the opposite sense to the motion induced by the
stabilizing Coriolis force (see, e.g. DP07). The frozen-in
electrons' motion which has destabilizing effect on the disk fluid
is in the same sense of the induced whistler circular motion of the
field lines (Balbus \& Terquem 2001, hereafter BT01). It was pointed
out by BT01 and DP07 that the drifting electrons impart a circularly
polarized component to the velocity response which damps the
Coriolis force and suppresses the stabilizing dynamical epicyclic
motion.

In the analysis of the diamagnetic effect on MRI, DP07 found that the magnetic field gradient
generated by the magnetization current produces a new unstable mode.
They also showed that the maximum growth rates and the parameter
spaces for the unstable modes depend strongly on the magnitude of
the magnetization and the magnetic field gradient it produces. In
their investigation, they assumed that the perpendicular velocity
$v_{\perp}$ does not vary in space around the fiducial radius R.
Therefore, they consider the effect of magnetic field gradient only.
Formally speaking, they dropped the third term on the right-hand
side of Eq. (6) containing $\nabla v_{\perp}$. In this study, we
will include the effects of both the magnetic field and
perpendicular velocity gradients for the first time.

\subsection{The geometry of disk-magnetosphere boundary}

\begin{figure}
\epsscale{1}
     \plotone{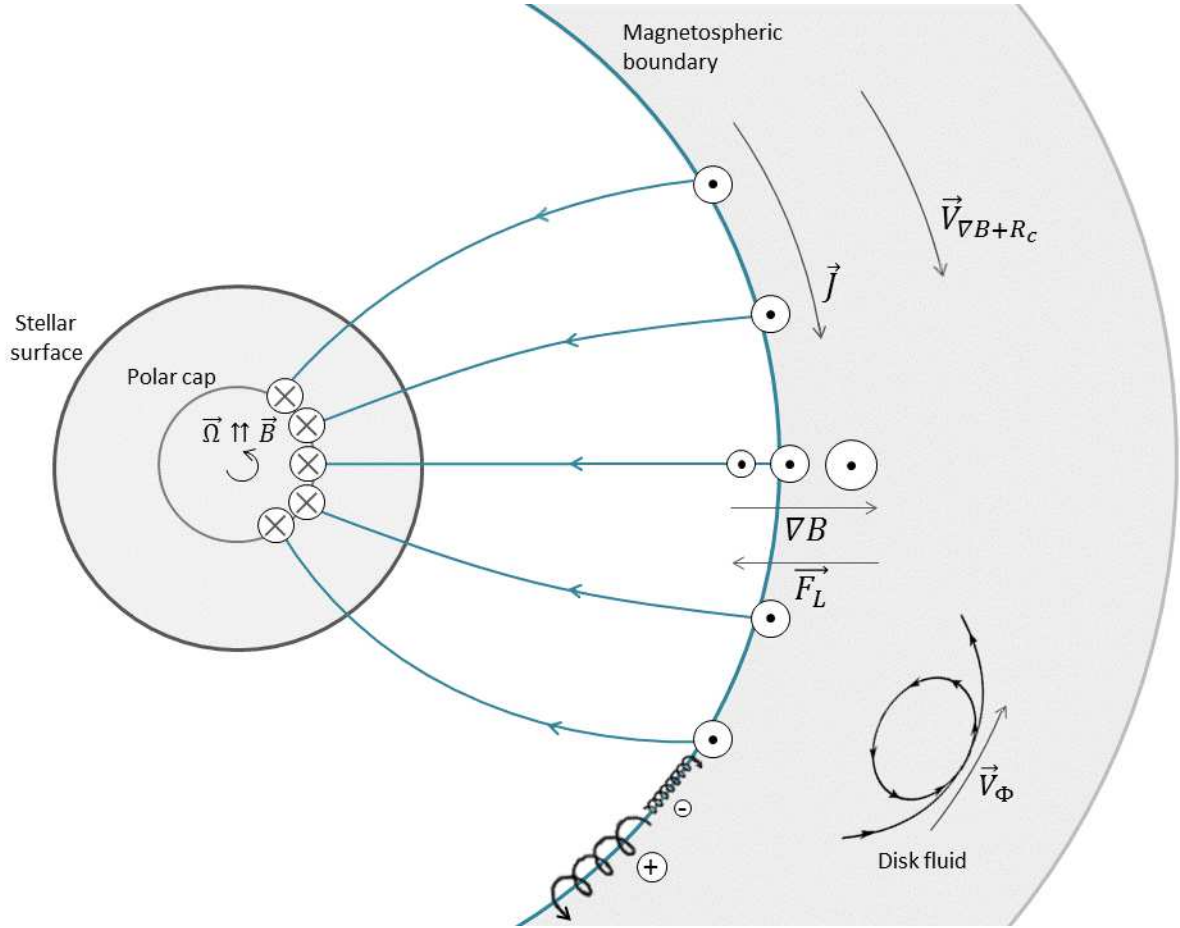}
    \caption{Top view of the border between the inner magnetosphere and
    the accretion disk (not to scale). Symbols $\odot$ and
    $\otimes$ represent the magnetic field vectors opposite and the same direction
    to the \emph{los} respectively. Cycloids at the inner radius of the disk represent the electron
    and proton trajectories. Drift motions of the trapped particles with a velocity
    $\vec{V}_{\nabla B+R_{\rm c}}$ generate diamagnetic current \textbf{J}. This current produces its own magnetic field in
    such a way as to produce a magnetic field gradient  $\nabla B$ at the border.  Generated
    magnetic pressure force combined with the tension force due to the curved field
    lines exert Lorentz force $\vec{F}_{\rm L}$ to the particles.  Epicyclic motion of the disk fluid
    is also shown at the bottom right of the figure.}
\end{figure}

\begin{figure}
\epsscale{1.1}
      \plotone{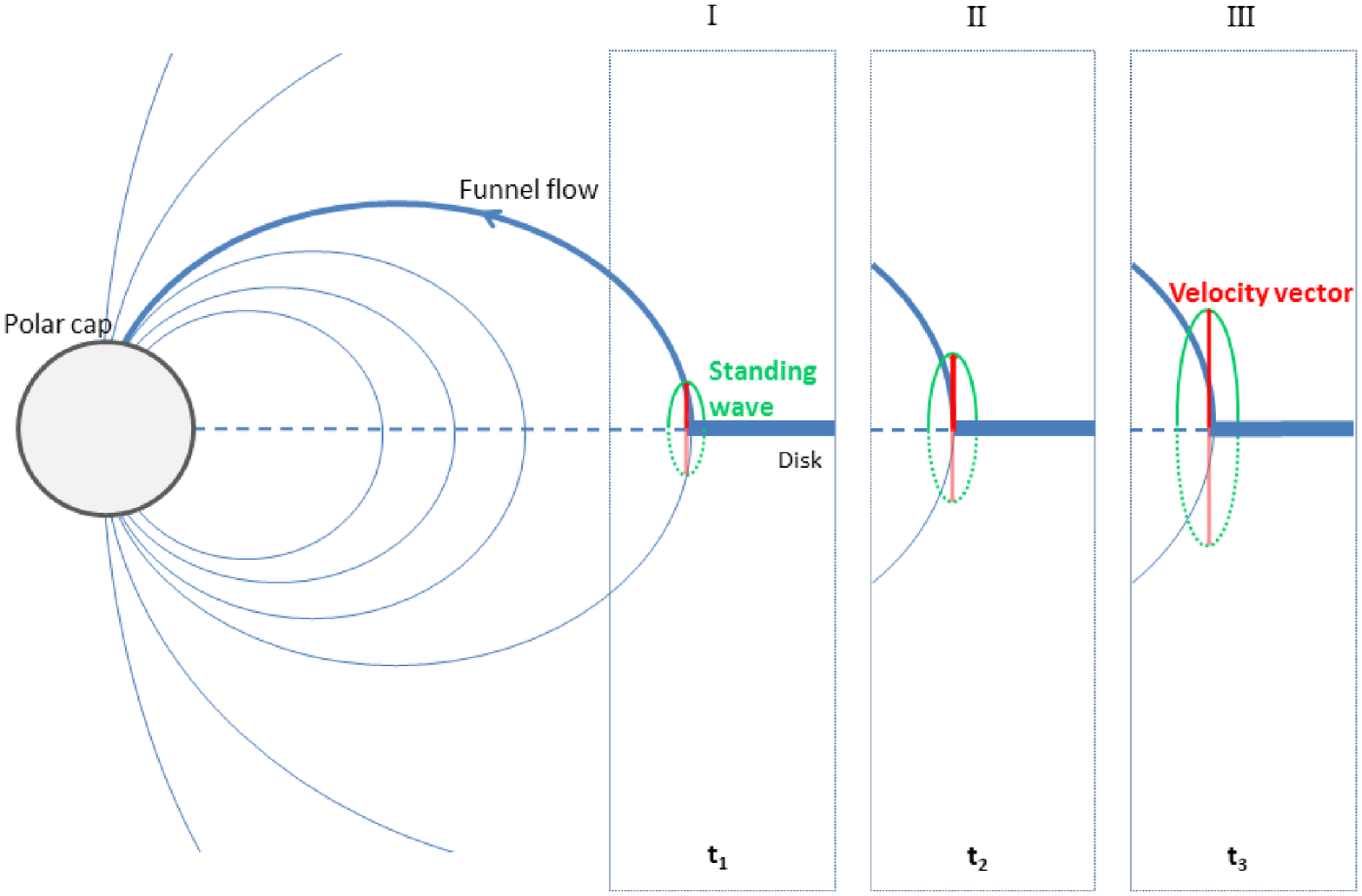}
    \caption{The inner boundary of the accretion disk is assumed to be
    co-rotating with the last closed field line of the magnetosphere.
    The amplitude of an unstable standing wave which is generated by
    \emph{diamagnetic current modifed MRI} at this border grows by time (see section 3) and
    pushes the disk fluid along the magnetic field lines with approximately
    zero pitch angle. As a result of this combination, i.e., gravitational
    pull by the central object and magnetic field guidance, funnel flow
    towards the magnetic polar regions becomes possible. Growing unstable
    wave amplitude is represented by a sinusoidal curve. As the unstable standing wave
    amplitude grows, disk fluid is channeled upward (continuous curve) and downward (dotted curve) directions
    towards the magnetic polar regions.}

\end{figure}

We assume a differentially rotating disk which manifests
itself in $\kappa$ (=4$\Omega^{2}$d$\Omega^{2}$/d $\ln$ R) epicyclic
frequency and the dimensionless parameter $\emph{T}$ (=d $\ln$
$\Omega^2$/d $\ln$ R) where $\Omega$ is the angular velocity
of the disk. Next, we also assume that the disk fluid
to be coupled to the weakly magnetized disk. These
two assumptions indicate that if an instability arises it is highly
likely to be MRI. We have taken neither self - gravity of the
disk nor the gravitational force of the CTTS into account.
Otherwise, one may argue the role of Rayleigh - Taylor instability.
The kinetic energy of the relative motions of drifting electrons and
positively charged particles is fed into a plasma wave of the
suitable phase velocity. The amplitude of this properly selected
wave may grow and result in instability (Somov 2006). We assume that
the last closed field lines co-rotate both with the star and the
inner boundary of the accretion disk. If this assumption is
justifiable then we may proceed as follows. Since the disk
fluid at the inner boundary is expected to be highly ionized and
transfer angular momentum outwards from the star, particles move
inward (BH91, BH98). They eventually become trapped in the
magnetosphere of the star and simultaneously acquire drift
velocities. Due to the curvature and the gradient of the magnetic
field, electrons and positively charged particles drift in the
opposite directions and thus produce a current flowing along the
inner boundary, just like the ring current that flows in the
terrestrial magnetosphere. This current generates its own magnetic
field and causes diamagnetism at the inner disk boundary.
The width of the region where this current flows is of the
order of the ions' average Larmor radius ($r_L=mv_{\perp}c/qB$)
which is found to be $200$ cm for CTTS. The net magnetic field
within that part of the magnetosphere that is neighboring the
diamagnetic current circuit is weaker than that of just outside the
current carrying circuit. Thus, a gradient of a magnetic field is
produced across the disk-magnetosphere boundary pointing
radially out from the star. If we assume that the magnetic
moment, ($\mu=mv_{\bot}^{2}/2B$) is conserved for the particles,
the particles$'$ perpendicular velocities, vary while they enter
into and exit from the higher magnitude field regions.

When the magnetic and rotation axes are not aligned,
the drifting disk fluid particles will spend half the
\emph{drift period} on the northern magnetic hemisphere and
the other half on the southern one. Let us try to depict the
magnetic course of the drifting particles. We assume that
positively charged particles, starting from the magnetic
equator, are drifting towards the magnetic northern latitudes.
They would encounter higher and higher magnetic field intensity
through a \emph{quarter-drift-period} (\emph{qdp}). The next
\emph{qdp} would be the time to descend towards the magnetic
equator. The drifting particles would encounter a similar
ascend and descend towards the southern magnetic pole at the
second half of the \emph{drift period}. During this periodic
ascending and descending from the magnetic polar regions, drifting
particles would encounter magnetic field gradients in the
azimuthal direction, in addition to the radial one generated
by the diamagnetic current. In this study, we will not consider
the non-aligned case and will put it off until a future study.

Let us consider a simple case wherein the rotation axis and
the magnetic moment axis are coincident, i.e., $\vec{\Omega}$
$\uparrow \uparrow$ $\vec{B}$ (see Fig. 1). In the \emph{top view}
outer circle represents the CTTS projected on a ($\emph{R}$, $\phi$)
plane in cylindrical coordinates; inner circle represents the
latitude of the last closed field lines on the star. The innermost circular arc is
the inner boundary of the disk; the symbols $\odot$ and $\otimes$
stand for the magnetic field vectors the directions of which are the
opposite and the same, respectively, to the \emph{los} of the
reader. Cycloid which is not drawn to the scale represents the
stabilizing epicyclic motions of the disk fluid. Diamagnetic current produced magnetic field
adds to the stellar field outside the circuit and subtracts at the inside
and thus brings about a magnetic field gradient which points in the
$\emph{R}$ direction. The Lorentz force due to the magnetic pressure
is in the negative $\emph{R}$ direction. At the same location, the
direction of the magnetic tension force is also in the negative
$\emph{R}$ direction and thus net force is strengthened.

If the drift velocity, \textbf{$V_{\nabla B+R_{\rm c}} \propto
R_{\rm c}\times B$} where $R_{c}$ is the radius of curvature of the magnetic field at the inner border of the disk, persists which in turn keeps the diamagnetic
current flow in the same sense as the induced whistler circular
motion then the diamagnetic effect destabilizes the disk just like
the whistler circular motion (DP07).

Diamagnetic-current-producing drifts may be described as
a self-organization process. It is argued that these kind of processes
keep their presence even at the turbulent stage (e.g. Fridman et al. 2006,
Prigogine \& Stengers 1984). disk-magnetosphere
interaction region may be defined as an open non-linear
magnetohydrodynamic system. This system can generate macroscopic
dissipative structures with various spatial and temporal scales by
an internal self organization like diamagnetic current. One may argue that this
current can maintain itself even at the onset of the
turbulence. But here, we make no claims as to the nonlinear regime
which is to be explored by numerical simulations. Our argument about
the diamagnetic currents is valid only in the linear regime.

Finally, let us imagine a slow and/or standing unstable wave
produced at the inner edge of the disk (see Fig. 2). In the next section we
will show that the solution of the dispersion relation reveals the
existence of an unstable standing wave. Therefore we deem it
necessary to visualize the physical process at the
disk-magnetosphere boundary by drawing the Fig. 2. The disk fluid is
forced by gravity which manifests itself by Keplerian velocity
profile, towards a CTTS and guided by the magnetic field. Now, the
growing amplitude of the unstable mode is highly likely to push the
disk matter parallel to the local magnetic field lines, i.e., with
an approximately zero pitch angle. This means that species of the
disk fluid, under this condition, will seek their respective mirror
points at the stronger parts of the magnetic field and eventually
hit the magnetic polar regions.

If the above assumptions we make are justifiable then we expect the
physical processes described above to take place. The
different drift velocities acquired by electrons and ions may bring
about the Hall currents as well. By taking into account all these
probabilities we set ourselves to the task of investigating the
possible role of MRI in producing slow or standing wave modes which may
guide the disk fluid along the magnetic field lines and thus cause
the precipitation of the plasma towards the magnetic poles of the
star.

\subsection{Equations}

The fundamental equations are mass conservation, the equation of
motion and the induction equation, given below, respectively:

\begin{equation}
\frac{\partial \rho }{\partial \, t} +\nabla \cdot \left(\rho \,
\textbf{\emph{v}}\right)=0
\end{equation}

\begin{equation}
\rho \frac{\partial \, \textbf{\emph{v}}}{\partial \, t} +\left(\rho
\textbf{\emph{v}}\cdot \nabla \right)\textbf{\emph{v}}=-\nabla
P+\frac{1}{c} \textbf{\emph{J}}\times \textbf{\emph{B}}
\end{equation}

\begin{equation}
\frac{\partial \textbf{\emph{B}}}{\partial t} =\nabla \times
\left[\emph{\textbf{v}}\times \textbf{\emph{B}}-\eta \frac{4\pi }{c}
\textbf{\emph{J}}-\frac{\textbf{\emph{J}}\times
\textbf{\emph{B}}}{en_{e} } \right].
\end{equation}

Current density (\emph{\textbf{J}}) which is given by Eq. (6) will
be substituted into Eqs. (8) and (9) in order to analyse
the influence of the diamagnetism on MRI. As a result, the equations
of modified momentum conservation and magnetic induction turn out to
be

\begin{equation}
\rho \frac{\partial \, \emph{\textbf{v}}}{\partial \, t} +\left(\rho
\emph{\textbf{v}}\cdot \nabla \right)\emph{\textbf{v}}=-\nabla
P+\frac{1}{4\pi } \left[{\rm (}1-\varepsilon {\rm )}\left(\nabla
\times \emph{\textbf{B}}\right)\times \emph{\textbf{B}}+2\varepsilon
\left(\frac{\nabla B}{B} \times\emph{\textbf{ B}}\right)\times
\emph{\textbf{B}}-2\varepsilon \left(\frac{\nabla {\rm v}_{\bot }
}{{\rm v}_{\bot } } \times \emph{\textbf{B}}\right)\times
\emph{\textbf{B}}\right]
\end{equation}

\begin{equation}
\frac{\partial \emph{\textbf{B}}}{\partial t} =\nabla \times
\left[\begin{array}{l} {\emph{\textbf{v}}\times
\emph{\textbf{B}}-\eta {\rm (}1-\varepsilon {\rm )}\nabla \times
\emph{\textbf{B}}-2\varepsilon \eta \frac{\nabla B}{B} \times
\emph{\textbf{B}}+2\varepsilon \eta \frac{\nabla {\rm v}_{\bot }
}{{\rm v}_{\bot } } \times \emph{\textbf{B}}-{\rm (}1-\varepsilon
{\rm )}\frac{c\left(\nabla \times \emph{\textbf{B}}\right)\times
\emph{\textbf{B}}}{4\pi en_{e} } } \\
\\
 {-2\varepsilon \frac{c}{4\pi en_{e} } \left(\frac{\nabla B}{B}
\times \emph{\textbf{B}}\right)\times \emph{\textbf{B}}+2\varepsilon
\frac{c}{4\pi en_{e} } \left(\frac{\nabla {\rm v}_{\bot } }{{\rm
v}_{\bot } } \times \emph{\textbf{B}}\right)\times
\emph{\textbf{B}}}
\end{array}\right].
\end{equation}

We consider the local stability of a differentially rotating disk
threaded by a vertical field with a gradient in the radial
direction, $\emph{\textbf{B}} = B(R)\hat{z}$. Therefore the gradient
of the magnetic field can be expressed by $\nabla B= (d B/d
R)\hat{\textbf{R}}$. The perpendicular velocity of the particles
also has a gradient in the radial direction, i.e. $\nabla v_{\bot}=
(d v_{\bot}/d R)\hat{\textbf{R}}$. We assume that finite resistivity
and Hall currents are both present. We shall work in the Boussinesq
limit. The Boussinesq approximation is frequently used in
descriptions of the nature of accretion disk transport. For
instance, BH98 argue that velocity field in accretion disks may be
taken as incompressible ($\nabla\cdot\textbf{u}=0$) for the
turbulent flows. They also warn the reader that the disk fluid
\textquotedblleft is not exactly but nearly\textquotedblright\,\,
incompressible. In detail, they say that
$(\nabla\cdot\textbf{u})^{2}$ is negligible compared to
$\mid\nabla\times \textbf{u}\mid^{2}$ and also
$\bigtriangledown(\bigtriangledown\cdot\textbf{u})$is negligible
compared to $\bigtriangledown^{2}\textbf{u}$, but
$P\bigtriangledown\cdot\textbf{u}$ is to be kept in the thermal
equation. This is how the Boussinesq approximation defined in the
context of accretion disks. We use standard cylindrical coordinates
(\emph{R},$\phi$,\emph{z}) with the origin at the disk center.
Finally, we assume that the perturbed quantities' variation in space
and time is like a plane wave, i.e., $\exp(ikz+\omega t)$, where
\emph{k} is the wave number perpendicular to the disk and
$\omega$ is the angular frequency. This form keeps the coefficients
of the dispersion relation real and a positive real root $\omega$
implies unstable exponential growth of the mode. Since we are
interested in plasma motion perpendicular to the disk plane,
 we investigate the instability of the mode with a wave vector perpendicular
to the disk. Under these circumstances, the linearized form of Eq.
(7), (10) and (11) are found as:

\begin{equation}
\omega \delta v_{R} -2\Omega \delta v_{\phi }^{}
-\left(1-\varepsilon \right)\frac{ik_{z} }{4\pi \rho } B_{z} \delta
B_{R} +\frac{1}{4\pi \rho } \left[\left(1+3\varepsilon \right)\nabla
B-4\varepsilon B_{z} \frac{\nabla \Omega }{\Omega } \right]\delta
B_{z} =0
\end{equation}

\begin{equation}
\omega \delta v_{\phi } +\frac{\kappa ^{2} }{2\Omega } \delta v_{R}
-\left(1-\varepsilon \right)\frac{ik_{z} }{4\pi \rho } B_{z} \delta
B_{\phi } =0
\end{equation}

\begin{equation}
\frac{ik_{z} \delta P}{\rho } -\frac{1}{4\pi \rho }
\left[\left(1+\varepsilon \right)\nabla B-2\varepsilon B_{z}
\frac{\nabla \Omega }{\Omega } \right]\delta B_{R} =0
\end{equation}

\begin{equation}
ik_{z} B_{z} \delta v_{R} -\left[\omega +\eta {\rm (}1-\varepsilon
{\rm )}k_{z}^{2} \right]\delta B_{R} -{\rm (}1-\varepsilon {\rm
)}\frac{c}{4\pi en_{e} } B_{z} k_{z}^{2} \delta B_{\phi } =0
\end{equation}

\begin{equation}
\begin{array}{l} {\left\{\frac{d\Omega }{d{\it ln}R} +\frac{c}{4\pi en_{e} } \left[{\rm (}1-\varepsilon {\rm )}k_{z}^{2} B_{z} {\rm +(}1+\varepsilon {\rm )}\nabla ^{2} B-2\varepsilon \left(B_{z} \frac{\nabla ^{2} \Omega }{\Omega } +\nabla B\frac{\nabla \Omega }{\Omega } -B_{z} {\rm (}\frac{\nabla \Omega }{\Omega } {\rm )}{\kern 1pt} ^{2} \right)\right]\right\}\delta B_{R}}\\
\\
{ -\left[\omega +\eta {\rm (}1-\varepsilon {\rm )}k_{z}^{{\rm 2}}
-2\varepsilon \eta \left(\frac{\nabla ^{2} B}{B} -{\rm
(}\frac{\nabla B}{B} {\rm )}^{2} -\frac{\nabla ^{2} \Omega }{\Omega
} +{\rm (}\frac{\nabla \Omega }{\Omega } {\rm )}{\kern 1pt} ^{2}
\right)\right]\delta B_{\phi } }\\
\\
 {+ik_{z} B_{z} \delta v_{\phi }+\frac{c}{4\pi en_{e} }
2\varepsilon \left[ ik_{z} \nabla B-ik_{z} B_{z} \frac{\nabla \Omega
}{\Omega } \right]\delta B_{z} =0}
\end{array}
\end{equation}

\begin{equation}
\begin{array}{l} {\nabla B\delta v_{R} +2\varepsilon \eta ik_{z} B_{z} \frac{\nabla
\Omega }{\Omega } \delta v_{\phi } -\frac{c}{4\pi en_{e} } \left[\,
{\rm (}1+\varepsilon {\rm )}ik_{z} \nabla B-2\varepsilon ik_{z}
B_{z} \frac{\nabla \Omega }{\Omega } \right]\delta B_{\phi }}\\
\\
{-\left[\omega +\eta {\rm (}1-\varepsilon {\rm )}k_{z}^{{\rm 2}}
+2\varepsilon \eta \left(ik_{z} \frac{\nabla B}{B} -\frac{\nabla
^{2} B}{B} {\rm +(}\frac{\nabla B}{B} {\rm )}^{2} -\frac{\nabla ^{2}
\Omega }{\Omega } +{\rm (}\frac{\nabla \Omega }{\Omega } {\rm
)}{\kern 1pt} ^{2} \right)\, \right]\delta B_{z} =0.} \end{array}
\end{equation}
where $\kappa=4\Omega^{2}d\Omega^{2}/d\ln R $ is the epicyclic
frequency and $\Omega$ is the angular velocity of the disk. The
linearized Eqs. (12)-(17) give a $\rm 5^{\rm th}$-order
dispersion relation that emerges after a very lengthy effort. This
describes five low-frequency modes. However, in the limit of zero
resistivity, the analysis is reduced to finding the roots of a
quartic. In this case, the dispersion relation in dimensionless form
is found as
\begin{equation}
s^4+\mathscr{C}_2s^2+\mathscr{C}_0=0
\end{equation}
where s = $\omega/\Omega$. The coefficients $\mathscr{C}_{2}$ and
$\mathscr{C}_{0}$ are found as follows:

\begin{equation}
\begin{array}{l} {\mathscr{C}_2={\widetilde{\kappa }}^2{\rm +2}{\rm X}\left({\rm 1-}
\varepsilon \right)+\frac{{\rm Y}}{4}\left({\rm 1-}\varepsilon
\right) \left[{\rm Y}\left({\rm 1-}\varepsilon \right){\rm
+T}\right]{\rm +} {{{\rm G}}^2\rm M}^{-2}_A\left[\left({\rm
1+3}\varepsilon \right)- \chi {\rm Y}\varepsilon \left({\rm
1+}\varepsilon \right)\right] }\\
{~~~~~~~~~{\rm +}\frac{{\rm 3}}{4} {\rm GT}\rm M^{-2}_A\chi {\rm
Y}\varepsilon \left({\rm 1+3}\varepsilon \right){\rm +}\frac{{{\rm
T}}^2}{8} \rm M^{-2}_A\chi {\rm Y}\varepsilon {\rm (8-9}\varepsilon
{\rm )}}
\end{array}
\end{equation}

\begin{equation}
\begin{array}{l} {\mathscr{C}_0=\rm G^2 \rm M^{-2}_A \{(1-\varepsilon)\left[\rm
X(1+3\varepsilon)+4\chi \rm X \varepsilon+\frac{\rm
Y^2}{2}(1-\varepsilon)\right]+(1+\varepsilon)\chi
\tilde{\kappa}^{2}\left[\rm Y\varepsilon-\frac{\rm
X}{4}(1+3\varepsilon)\right]\}}\\
{~~~~~~~~~~\rm +GTM^{-2}_A\left\{\begin{array}{l}{\frac{\rm \chi}{2}
\tilde{\kappa}^{2}(\rm X-\rm Y)\varepsilon (1+\varepsilon)-2(3\chi-2)
\varepsilon(1-\varepsilon)}\\
{-\frac{\rm \chi}{8}\rm XT\varepsilon(7+5\varepsilon)+\frac{\rm Y^2}{4}
\varepsilon^2(1-\varepsilon)^2
+\frac{\rm \chi}{2}\rm XG(1+3\varepsilon)(1+\varepsilon)}\\
{+\rm M^{-2}_A \chi \rm Y\left[\frac{\rm GT}{4}\varepsilon^2(1+3\varepsilon)
-2\varepsilon(1-\varepsilon)-\frac{\rm T^2}{8}\varepsilon^2
(11+3\varepsilon)-\frac{\rm G^2}{2}\varepsilon(1+3\varepsilon)\right]}\end{array}\right\}}\\
~~~~~~~~~~~{-\rm T^2\rm M^{-2}_A\left\{\begin{array}{l}{\frac{\rm \chi}{8}\rm Y
\tilde{\kappa}^{2}\varepsilon (1+3\varepsilon)+\frac{\chi}{2}\rm X\varepsilon
\left[(1-\varepsilon)-\tilde{\kappa}^{2}\varepsilon+\rm T\varepsilon\right]}\\
{+\frac{\rm \chi}{4}\rm M^{-2}_A\varepsilon^2\left[\rm T^2\varepsilon^2+2(1-\varepsilon)\right]
-\frac{\rm Y^2}{8}\varepsilon^2(1-\varepsilon))}
\end{array}\right\}}\\
~~~~~~~~~~~{+\left[\rm T(1-\varepsilon)+\rm Y(1-\varepsilon)^2+\rm X(1-\varepsilon)^2\right]
(\frac{\tilde{\kappa}^{2}\rm Y}{4}+\rm X)}.
\end{array}
\end{equation}

Here the dimensionless parameters are defined as s =
$\omega/\Omega$, X = $(kv_{A}/\Omega)^{2}$, Y =
$(kv_{H}/\Omega)^{2}$, $\tilde{\kappa}= \kappa/\Omega$, G = d ln B/d
ln R, T = d ln $\Omega^2$/d ln R and $\chi \equiv
v_{H}^{2}/v_{A}^{2}$. The Hall and the Alfv\'{e}n speeds are
defined as $v_{\rm H}^{2}=\Omega Bc/2\pi en_{\rm e}$ and $v_{\rm
A}^{2}=B^{2}/4\pi \rho$. The Alfv\'{e}n Mach number of Keplerian
(orbital) motion $M_{A} = v_{\phi}/v_{A}$ measures the relative
strength between the kinetic energy and the magnetic energy. We can
rewrite the Alfv\'{e}n Mach number in terms of $\varepsilon$ as
$M^{2}_{A} = 3\varepsilon$.

In this case, the dispersion relation gives two fast and two slow
modes which are labeled depending on the magnitude of their phase
velocities. Besides, in Figs. 3-5 which are given in the next
section, ridges of the growth rates correspond to the sites (in X,Y)
of standing waves, i.e., the solution of dispersion relation gives
real $\omega$. No harm in repeating that we assumed the
phasor factor of the waves as $\exp(ikz+\omega t)$. Therefore, the
slow mode turns into non-propagating, i.e. standing mode in the
(X,Y) regions of instability, since $\omega_i=0$. disk fluid will be
pushed up and down along the magnetic field while the magnitude of
the standing wave grows through instability. This, we believe, is
the way to lift the material from the disk into funnel flow. The
disk material gains momentum in this way and
enters into the magnetic field with almost zero pitch angle
and be guided towards the magnetic poles of the star under
gravitational force.

We seek the solution at a fiducial radius (R).
In this investigation, the fiducial radius corresponds to the close neighbourhood of the
\emph{Alfv\'{e}n radius} where plasma pressure is slightly bigger than the magnetic pressure.

Eq. (18) is the general form of the dispersion relation,
wherein the gradient in the magnetic field ($\nabla B$) and
the perpendicular velocity ($\nabla v_{\bot}$) are both considered.
This amounts to saying that the magnetization currents are
persistent and strong in the disk-magnetosphere boundary. Therefore,
the long-lasting currents can produce a gradient in the magnetic
field and perpendicular velocity in turn. In the next section, we
will compare our results for three cases: \emph{i)} $\nabla B=0,
\nabla v_{\bot}=0$; \emph{ii)} $\nabla B\neq0, \nabla v_{\bot}=0$;
\emph{iii)} $\nabla B\neq0, \nabla v_{\bot}\neq0$.

\section{Numerical Growth Rates of the Unstable Mode}

The graphical solutions of Eq. (18), the numerical growth rates (s),
are thus shown in (X,Y) plane in Figs. (3)-(5). Here we present the
results for three cases. First we assume that
the magnetization current is not persistent but fluctuates, so that
the magnetic field and perpendicular velocity gradients produced at
the disk-magnetosphere boundary can be neglected. Formally speaking,
the second and the third term on the right-hand side of Eq. (6) will
be dropped, i.e. $\nabla B=0$ and $\nabla v_{\bot}=0$. Then, we
consider a long-lasting current and take the magnetic field gradient
into account, but the perpendicular velocity still remains constant
in space around the fiducial radius R, i.e. $\nabla B\neq0$ and
$\nabla v_{\bot}=0$. Finally, we consider the effects of both the
magnetic field and perpendicular velocity gradients, i.e. $\nabla
B\neq0$ and $\nabla v_{\bot}\neq0$. Therefore we can compare the
results for these three cases.

\begin{table*}
\caption{The maximum numerical growth rates found from Solution I and II.} 
\begin{tabular}{llcccc}
&& \multicolumn{4}{c}{$s_{\rm m}$}\\
\hline
&&\multicolumn{2}{c}{$\varepsilon=0.1$} & \multicolumn{2}{c}{$\varepsilon=0.5$}\\
& & FUR & SUR & FUR & SUR \\
\hline 
&   &   &   &   & \\
Solution I &G=0.0&0.75&-&0.75&-\\
&   &   &   &   & \\
Solution II&G=0.1&0.78&0.22&0.79&0.10\\
  &G=0.5&1.37&1.02&1.58&0.82\\
&G=1.0&2.43&2.21&4.42&2.65\\

\hline
\end{tabular}

\medskip
The Solution I (G=0, T=0) reveals one unstable region. The Solution
II (G$\neq$0,T=0) reveals two unstable regions. We label them as FUR
(First Unstable Region), and SUR (Second Unstable Region).
\end{table*}

\begin{table*}
\caption{The maximum growth rates for Solution III.}
\begin{tabular}{llcc}

&&\multicolumn{2}{c}{$s_{\rm m}$}\\
\hline
&& FUR & SUR  \\
\hline
Solution III&G=0.1, $\varepsilon=0.1$, T=-3 &1.41&1.52\\
  &G=0.5, $\varepsilon=0.4$, T=-3.5&-&14.78\\
&G=1.0, $\varepsilon=0.7$, T=-4&-&29.13\\
\hline
\end{tabular}
\end{table*}

BT01 discussed the Hall effect on MRI in protostellar disks
from a dynamical point of view. Although they focused on
protostellar disks only, they claimed that their results are broadly
applicable. They analyzed the relative importance of the Hall term
both for gases with a low-ionization fraction and for fully ionized
plasma. One of the important conclusions of their analysis is that
the temperature and density regimes of ionized accretion disks imply
that the Hall effect cannot be ignored. Bearing this analysis in
mind, we choose to take the Hall effect into account.
We can estimate the Hall parameter for TT disks as follows: TT disks are
expected to be truncated at distances of several stellar radii from
the star (Johns-Krull 2007). Bouvier et al. (2007) found the truncation
radius where the stellar magnetic field starts to control the motion
of the accreting plasma as about 7 stellar radii for $B_*$= 1 kG.
Then at $R_{\rm M}=7 R_{*} $ the magnetic field strength is found as
$B_{\rm M}=B_{*}(R_{*}/R_{\rm M})^{3}\approx 2.9$ G. Typical
midplane gas density in TT disks is given as $\rho_g=10^{-9} \rm
gcm^{-3}$ (Alexander 2008). Glassgold et al. (2007) found the
electron number density as $10^5 \rm cm^{-3}$ from Neon fine
structure line emission of a TT disk. For a typical period $P \sim
8^{\rm d}$, the $\chi$ value is found as 4. We will use this value for the Hall
parameter in our numerical solutions.

\subsection{Growth rates in case of $\nabla B=0$ and $\nabla
v_{\bot}=0$}

\begin{figure*}
  \begin{center}
    \begin{tabular}{cc}
     \centering
      {\includegraphics[angle=270,scale=.6]{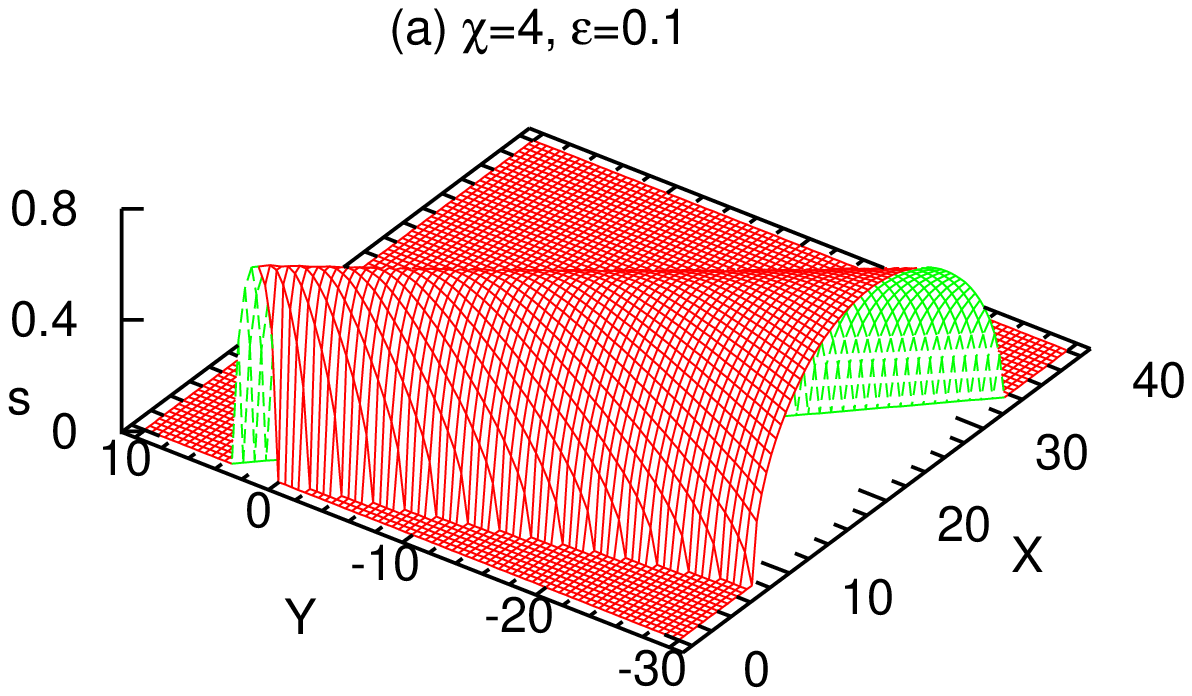}}&
      {\includegraphics[angle=270,scale=.6]{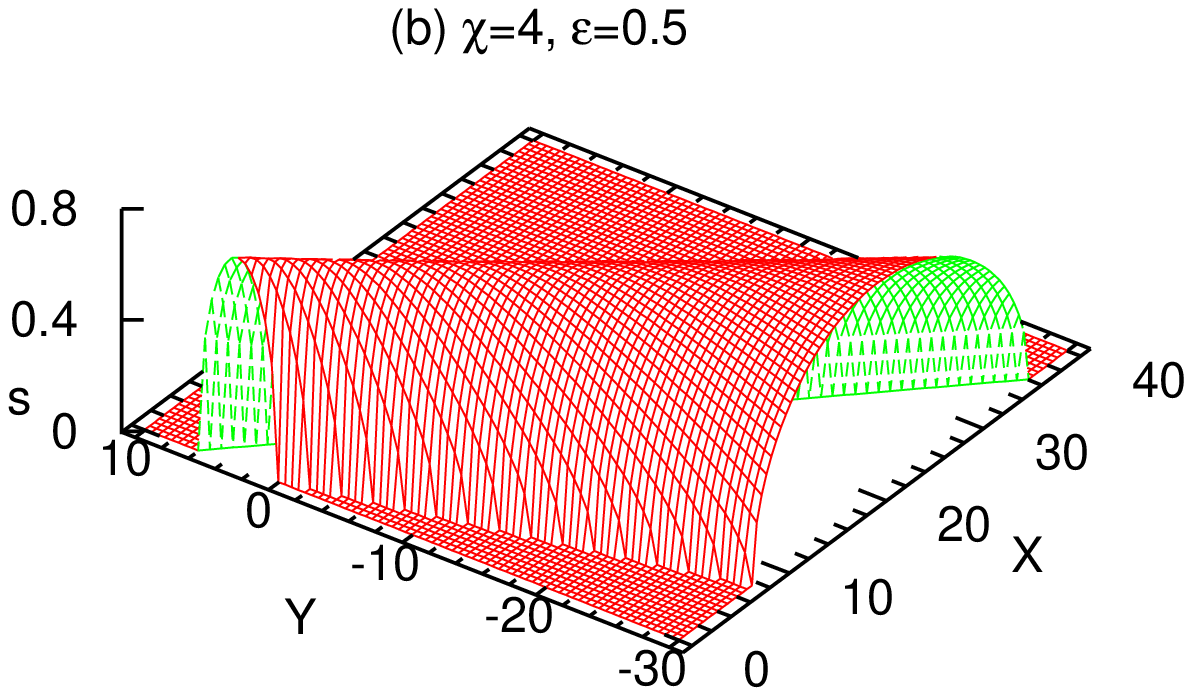}}\\
          \end{tabular}
    \caption{Growth rates found from Solution I. Only regions of instability are shown, with
the height proportional to the growth rate. The regions of
instability are seen as \textquotedblleft ridges\textquotedblright
above the X,Y plane (see text for definitions). The maximum growth
rate of the \textquotedblleft ridge\textquotedblright is $ 0.75$ for
both cases: (a) weak magnetization $(\varepsilon=0.1)$; (b) strong
magnetization $(\varepsilon=0.5)$. However, the unstable region
slightly widens with increasing $\varepsilon$.}
     \end{center}
\end{figure*}

As we mentioned above, in our first solution (Solution I) we neglect
the gradients of the magnetic field and the perpendicular
velocity generated by the magnetization current at the
disk-magnetosphere boundary. We assume a uniform vertical magnetic
field, $\textbf{B}=B\hat{z}$.

The graphical solutions are given
in Fig. 3. In order to see the effect of
magnetization on growth rates, we plot our graphs for a weak
$(\varepsilon=0.1)$ and strong $(\varepsilon=0.5)$ magnetization.
For both cases, the maximum value of the growth rate is 0.75 and
independent of the value of $\varepsilon$. However, it is apparent
that the region of instability which is seen as a ridge, slightly
widens when we increase the value of $\varepsilon$ (see Fig.
3b).

We should mention that when $\varepsilon=0$, the dispersion
relation is reduced to Eq. (57) of BT01 and the graphs of Solution I
are clearly similar to that of BT01, as they should be.

\begin{figure*}
  \begin{center}
    \begin{tabular}{ll}
      {\includegraphics[angle=270,scale=.6]{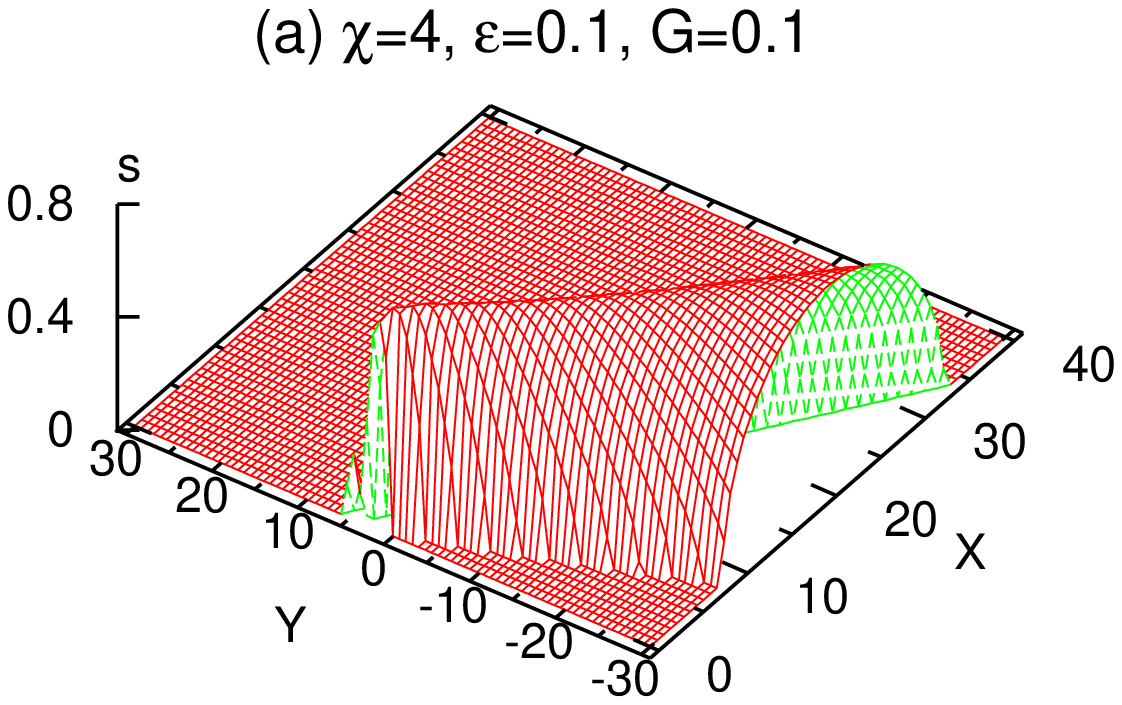}}&{\includegraphics[angle=270,scale=.6]{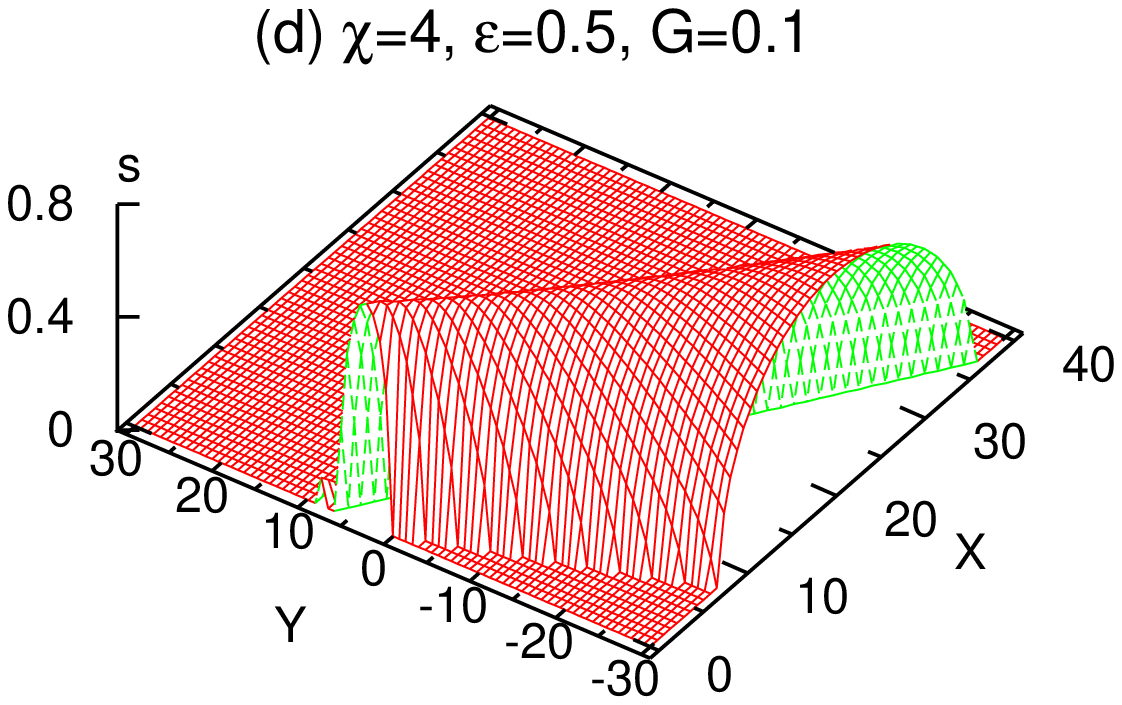}}\\
      {\includegraphics[angle=270,scale=.6]{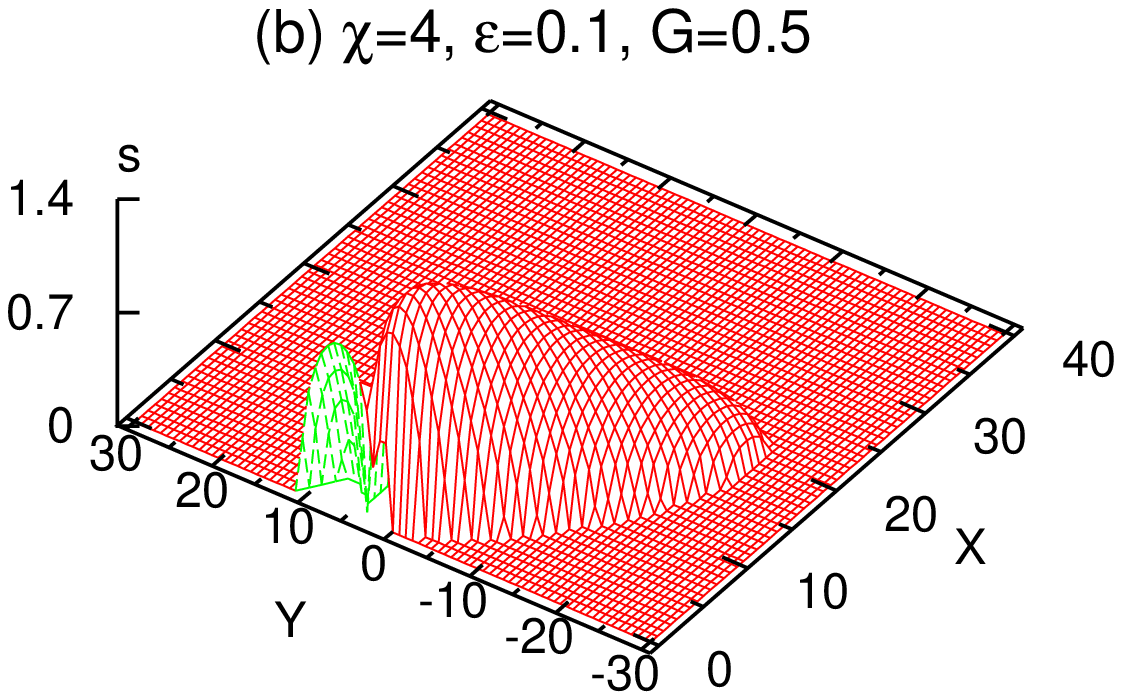}}&
      {\includegraphics[angle=270,scale=.6]{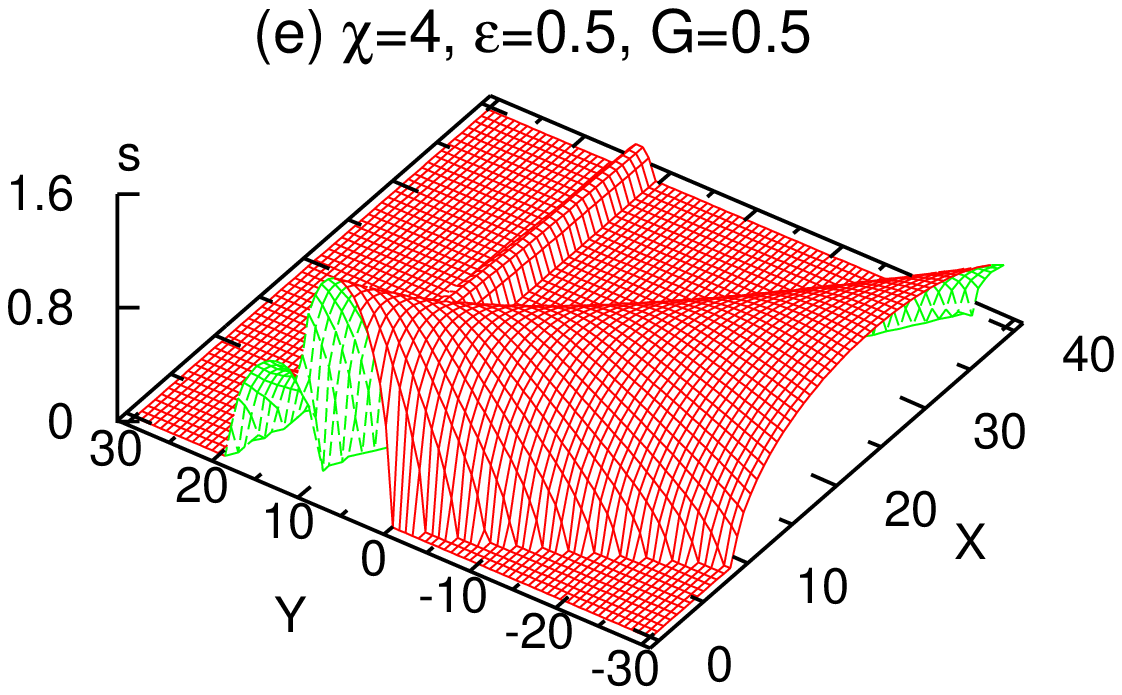}}\\
      {\includegraphics[angle=270,scale=.6]{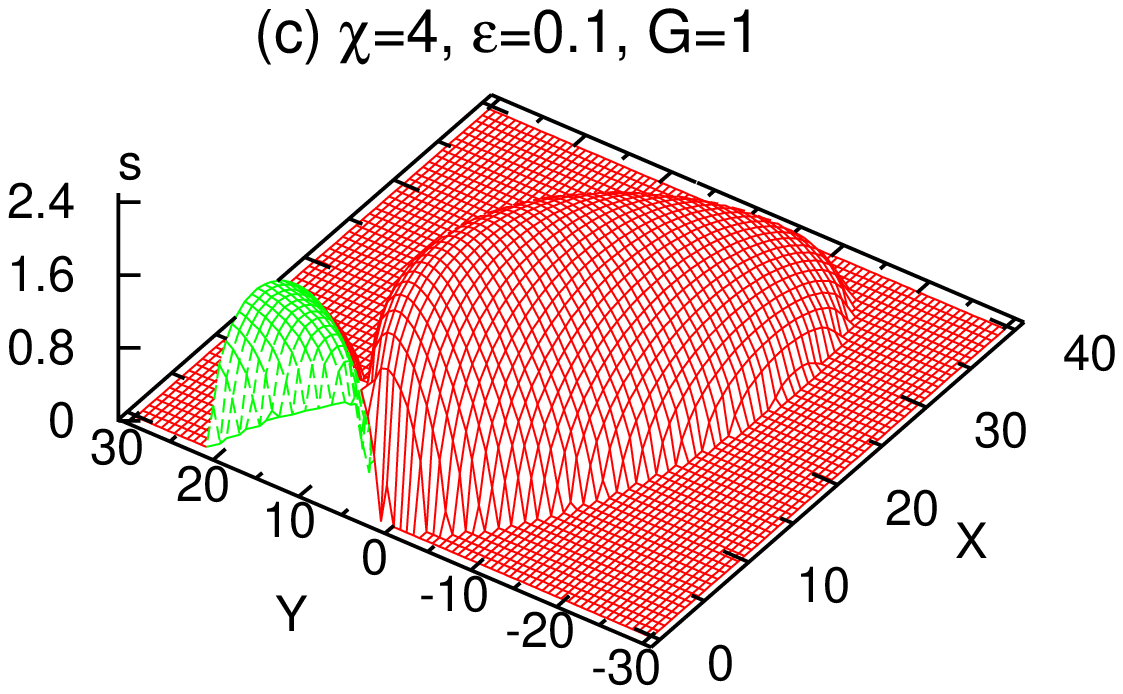}}&{\includegraphics[angle=270,scale=.6]{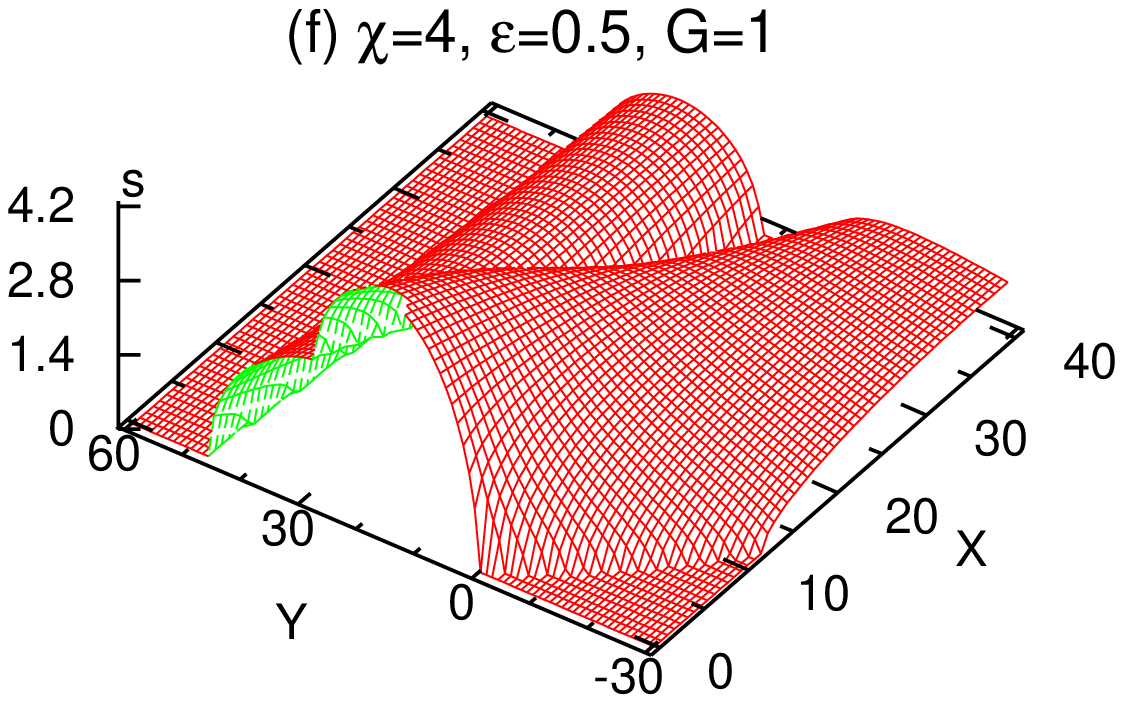}}\\
          \end{tabular}
    \caption{Growth rates found from Solution II for TT disks.
    In order to see the effect of $\nabla B$ on growth rates, we keep
    $\varepsilon$ constant and change the value of G from
    arbitrarily chosen values of 0.1, 0.5 and 1 (see text for definitions).
    Left panels are for weak magnetization $(\varepsilon=0.1)$;
the right panels are for strong magnetization $(\varepsilon=0.5)$.
A new unstable mode comes into existence with the inclusion of
$\nabla B$. See Table 1 for maximum values of the growth rates
($s_{\rm m}$).}
     \end{center}
\end{figure*}

\begin{figure*}
  \begin{center}
    \begin{tabular}{lll}
            {\includegraphics[angle=270,scale=.4]{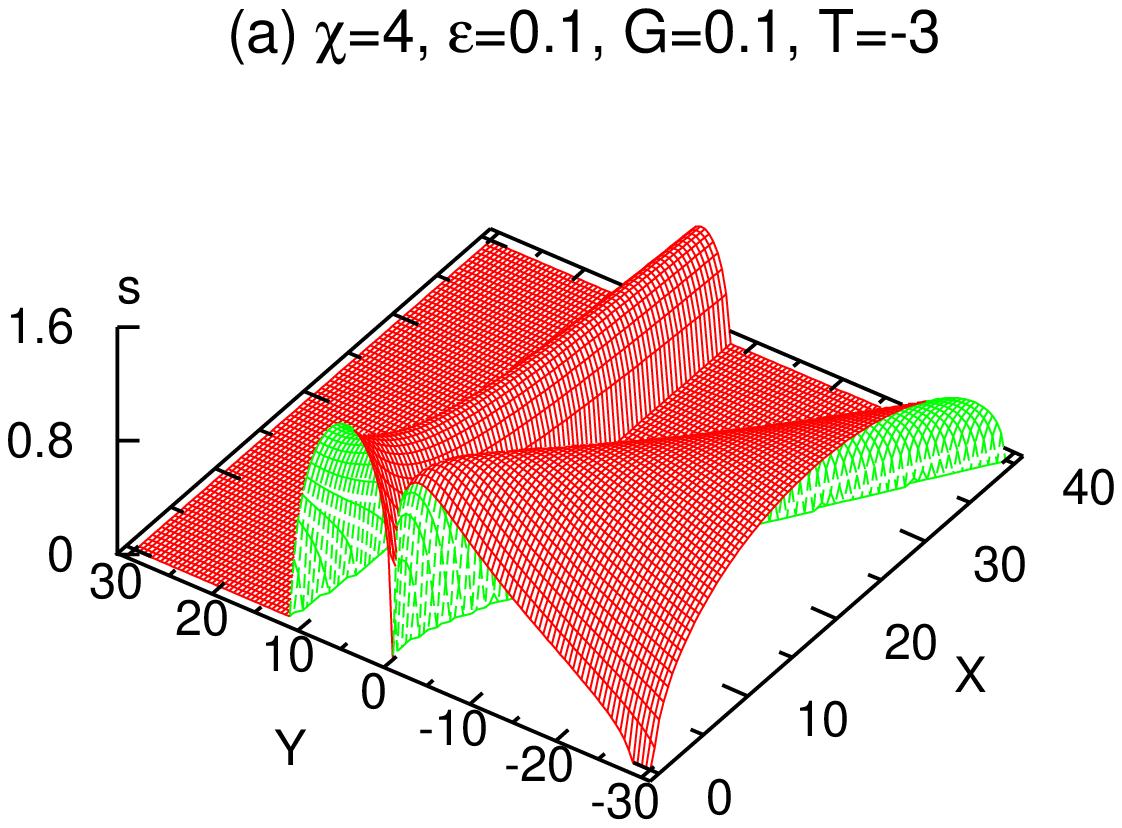}}&
      {\includegraphics[angle=270,scale=.4]{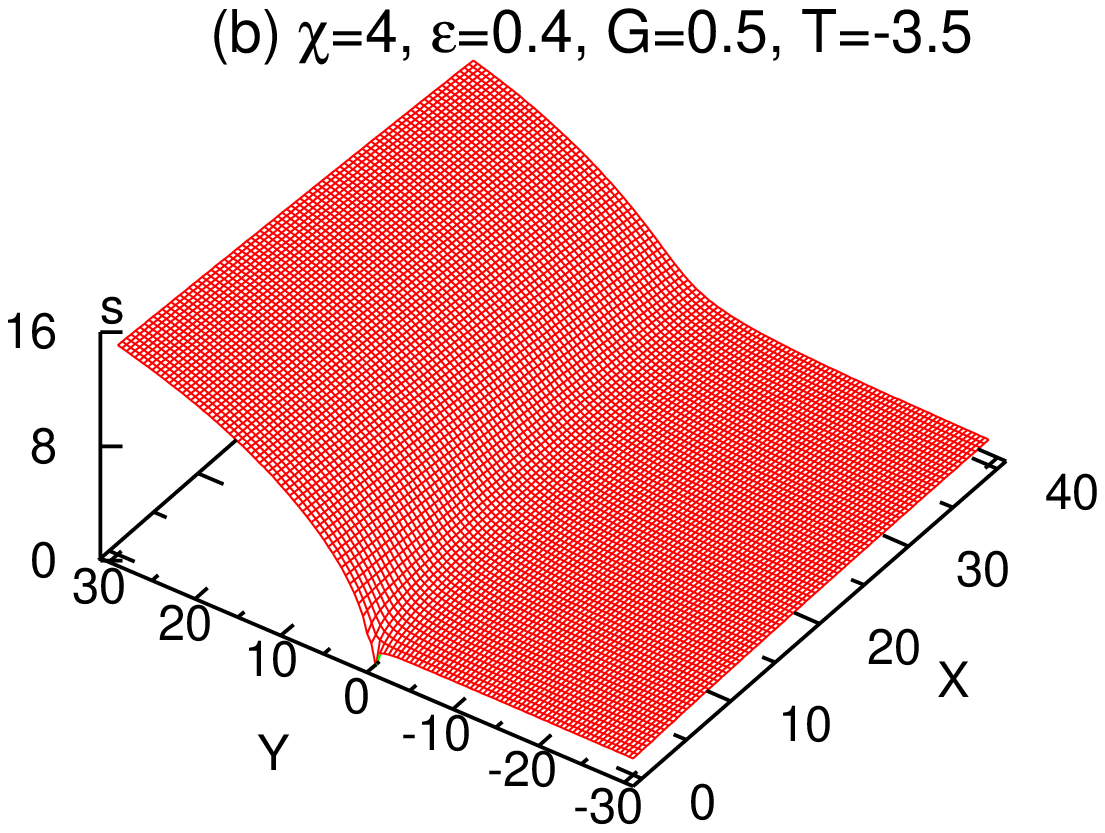}}&{\includegraphics[angle=270,scale=.4]{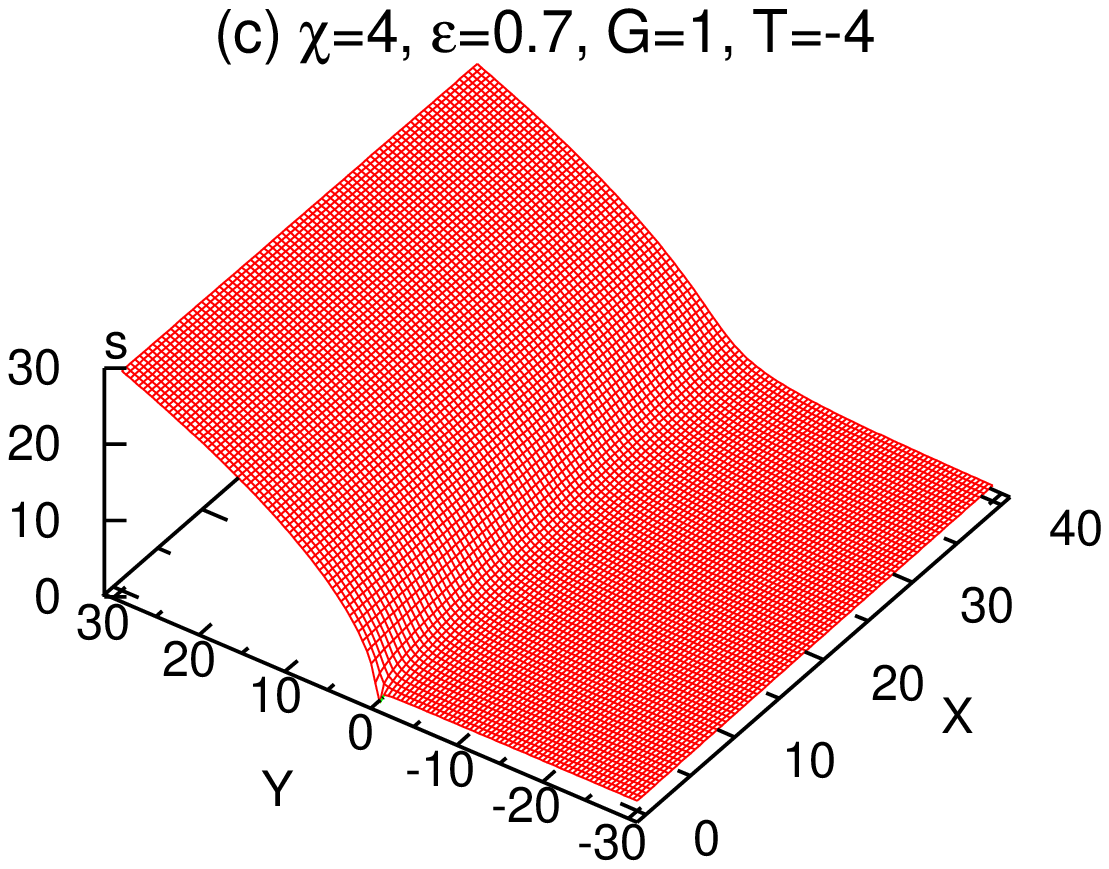}}\\
          \end{tabular}
    \caption{Growth rates found from Solution III. With the inclusion of $\nabla
v_{\bot}$ the maximum values of the growth rates increase and the
unstable regions widen in the X,Y plane. See Table 2 for maximum values of the growth rates. }
     \end{center}
\end{figure*}

\subsection{Growth rates in case of $\nabla B\neq0$ and $\nabla
v_{\bot}=0$}

In our second solution (Solution II), we consider a strong,
long-lasting magnetization current that can produce a gradient in
the magnetic field at the disk-magnetosphere boundary. The
equilibrium vertical magnetic field is assumed to vary in the radial
direction, i.e. $\emph{\textbf{B}}=B(R)\hat{\textbf{z}}$. Therefore
the gradient of the magnetic field can be expressed by $\nabla B= (d
B/d R)\hat{\textbf{R}}$. We still assume that the perpendicular
velocity does not vary in space at the disk-magnetosphere boundary
(i.e. $\nabla v_{\bot}=0$).

The results of Solution II are presented in Fig. 4. A new
unstable region comes into existence with the inclusion of $\nabla
B$. From now on, we refer to this new region of instability as the
\emph{second unstable region} (SUR) and the one which was found in
Solution I as the \emph{first unstable region} (FUR). We again
present the roots for a weak (see the left panel in Fig. 4)
and strong (see the right panel in Fig. 4) magnetization
respectively. In order to see the effect of $\nabla B$ on growth
rate, we keep $\varepsilon$ constant and change the value of G from
arbitrarily chosen values of 0.1, 0.5 and 1. In Figs 4a, b and c, we
keep $\varepsilon=0.1$ and increase the value of G. It is apparent
that when the value of G increases, the SUR becomes wider and the
maximum value of the growth rates becomes higher. For G=1, the
maximum growth rate which is included in the SUR, turns out to be
2.43 (see Fig. 4c). The maximum growth rates of the SUR and the FUR
are listed in Table 1 for different values of $\varepsilon$ and G.
In Figs 4d, e and f we keep $\varepsilon=0.5$ and increase
the value of G again. When we compare the results for weak
($\varepsilon=0.1$) and strong ($\varepsilon=0.5$) magnetization, we
see that the maximum values of the growth rates turn out to be
higher in the presence of strong magnetization. The maximum growth
rate reaches a value of 4.42 for G=1 and for $\varepsilon=0.5$. The
unstable regions (both FUR and SUR) again spread over a larger space
in the X,Y plane for $\varepsilon=0.5$ than they do for
$\varepsilon=0.1$.

We should mention that the results of Solution II are clearly
similar to that of DP07, since the inclusion of magnetic field gradient
is the same case considered in DP07. However, the maximum values of the
growth rates are found to be slightly lower than that of DP07. This is
highly probably a result of the Hall effect. The low value of the Hall parameter
decreases the growth rates.

\subsection{Growth rates in case of $\nabla B\neq0$ and $\nabla
v_{\bot}\neq0$}

In the previous solution, we assumed that the magnetization
current produces a gradient in magnetic field at the
disk-magnetosphere boundary. If the magnetic moment
($\mu=mv_{\bot}^{2}/2B$), is conserved for drifting
particles, the perpendicular velocity of the particles vary while
they enter into and exit from the higher magnetic field regions.
Therefore, a gradient in particles$'$ perpendicular velocity is
produced. In Solution III, we shall not omit the third term which
includes $\nabla v_{\bot}$ in Eq. (6). The graphical solutions
(Solution III) are presented in Fig. 5.
Fig. 5a shows the results for Keplerian rotation. Therefore
T = d ln $\Omega^{2}$/d ln R=-3. The inclusion of the term
containing $\nabla v_{\bot}$ both increases the maximum values of
the growth rates and widens the unstable regions in the X,Y plane. If we compare Fig. 4a and 5a, we
see that the maximum value of the FUR increases from 0.78 (see
Fig.3a) to 1.41, and also the maximum value of the SUR increases from
0.10 to 1.52 with the inclusion of the $\nabla v_{\bot}$. In Figs
5b and 5c, we investigate the case of deviation from Keplerian
rotation as a result of gradient in magnetic field and the
perpendicular velocities. If the gradient in magnetic field strength
and perpendicular velocity cause a change in Keplerian velocity
profile of the inner disk, T should deviate from the value of \textquotedblleft$-3$\textquotedblright.
Since the values of $\epsilon$, G and T are proportional to each
other, we increase the value of $\epsilon$ and G with increasing T
in Figs 5b and 5c. It is clearly seen that both the maximum values
of the growth rates become higher and the regions of instability
become wider with increasing $\epsilon$, G and T (see Table 2).

\section{Conclusion}
We investigated the stability of a disk around CTTS by
taking the diamagnetic effect into account. Plasma entry into the
magnetosphere was explained by various types of plasma instabilities
(e.g. the Rayleigh-Taylor instability, Kelvin-Helmholtz
instability), diffusion and magnetic reconnection processes, and
loss-cone mechanism in previous investigations. In this
study, we investigated the perpendicular unstable mode produced by
MRI which may be an alternative mechanism in guiding the plasma
above the disk at the magnetospheric boundary. The main conclusions
of our study can be summarized as follows:

\begin{enumerate}
  \item MRI produces an unstable mode which can raise the plasma
  from the disk towards the vertical magnetic field lines. The
  diamagnetic effect modifies the growth rate and the wavelength
  range of the unstable mode. In the simplest case, when the field and velocity
  gradients produced by magnetization are not included, the region of instability
  widens in the (X,Y) plane with increasing
  magnetization parameter ($\epsilon$).
  \item A new unstable region comes into existence with the
  inclusion of the gradient in the magnetic field (G). The maximum growth
  rate of the unstable mode, $s_{m}$, depends more strongly on G than
  on $\epsilon$. When we keep $\epsilon$ constant and change G, the maximum growth rate of the unstable mode
  increases with increasing values of G. On the other hand, the
  unstable region widens for higher values of $\epsilon$.
  \item The inclusion of the perpendicular velocity gradient
  in addition to the field gradient, increases the maximum
  growth rates and widens the regions of the instability. Moreover,
  the maximum value of the growth
    rates increase with increasing gradient in perpendicular
    velocity. As a result, MRI becomes more powerful when we include the gradients in magnetic field and perpendicular velocity.
\end{enumerate}

In the Hall-dominated regime electrons are frozen-into the magnetic field
but ions are coupled to the neutrals. At the inner radius of the disk
this situation enhances the diamagnetic current density, in turn,
sharpens the magnetic field gradient. The modification to the growth rate
by the magnetic field gradient may be seen in Figs. 3-5. The higher
the magnitude of the gradient the greater the growth rate. In this respect,
Wardle \& Salmeron (2012) investigated the effect of the Hall diffusion on
the stability of the Keplerian disk. They argue that in PPDs Ohm and ambipolar
diffusion have a stabilizing role while the Hall effect either stabilizes or
destabilizes the disk depending on the orientations of the magnetic field
and the rotation axes. And they warn the reader that small dust grains may
remove the electrons through recombination and the  MRI-active column density
reduces and MRI becomes irrelevant in PPDs.

As a result of the growing amplitude of the slow and/or
standing waves produced in the disk-magnetosphere boundary of the
magnetized star, we may expect that the disk fluid can be lifted
towards the magnetic field lines with an approximately zero pitch
angle. Zero pitch angle means that the plasma particles will seek
their magnetic mirrors at the regions of higher magnetic field. It
is highly probable that the particles will not be reflected until
they reach the magnetic polar regions. Therefore, they will
eventually hit the polar caps. The amplitude of the wave and the
velocity of the disk fluid along the magnetic field lines determine
the effectiveness of the lifting process. But this may be a subject
of a study of non-linear regimes.

Romanova et al. (2011) stress on the fact that ``the
turbulence in the disk is initiated and supported by the
magneto-rotational instability". This means that MRI producing
conditions, i.e., diamagnetic current etc. are continuous.
Turbulence should not be considered as noise or disorder. At the
macroscopic scale it may appear as chaotic but at microscopic scale
turbulence reveals itself as highly organized. Drifting charged
particles at the closed field lines of the magnetosphere move in a
coherent way so that the diamagnetic current persists within the
multiple time and length scales of turbulence. We may qualitatively
argue that diamagnetic current and the magnetic field it produces
may bring about a magnetic field gradient which in return triggers
MRI and causes the laminar flow go turbulent. We should emphasize
that in this study we work in the linear regime not in the
non-linear one. So that we are not in a position to claim that the
MRI will lead to turbulence at the non-linear regime.

Above all, it is yet to be seen as to whether the instability grows
into turbulence. This requires a numerical simulation which will be
the subject of a future study.



\acknowledgments
Our special thanks go to A. R. King and R. Lovelace for helpful
suggestions and B. Kalomeni and G. James for reading the manuscript. SD appreciates
the Theoretical Astrophysics Group for their hospitality. We dedicate
this study to the honorable scientific and organizational effort given
to the Turkish Astronomy by Prof. Dr. Zeki Aslan. This work
is supported by Turkish Academy of Sciences (T\"{U}BA) Doctoral
Fellowship. This study is a part of PhD project of SD.

\clearpage

\end{document}